\documentclass[twocolumn]{aastex631}

\usepackage{graphicx}
\usepackage{amsmath}
\usepackage{hyperref}

\shorttitle{Statistical Validation of TOI-7701.01}
\shortauthors{Escolà-Rodrigo}

\begin{document}

\title{Comprehensive Statistical Validation of TOI-7701.01: A Sub-Saturn Companion at the Giant Planet Boundary}

\author[0009-0007-1479-2386]{Biel Escolà-Rodrigo}
\affiliation{Independent Researcher}
\correspondingauthor{Biel Escolà-Rodrigo}
\email{escola.rodrigo.astro@gmail.com}

\begin{abstract}
We present the formal statistical validation of TOI-7701.01 (TIC 122522333), a sub-Saturn transiting companion candidate orbiting a bright F-type subgiant host star ($V=11.09$). Initially detected via an automated machine-learning transit survey by Salinas et al. (2025), the planetary nature of this signal remained unclear. We implement an independent vetting and statistical pipeline leveraging multi-sector space kinematics and the \texttt{triceratops} Bayesian framework to compute false positive scenarios. A key methodological finding is the structural convergence of the companion's physical scale; despite utilizing un-detrended Simple Aperture Photometry (SAP) with a raw depth of $2417$ ppm to preserve field dilution metrics, the \texttt{triceratops} MCMC engine naturally converges on a physical radius of $7.86\,R_\oplus$ under the dominant Target Planet scenario ($\mathcal{P}_{\rm TP} = 78.1\%$). This independent statistical derivation matches the geometric radius of $8.07\,R_\oplus$ inferred directly from the instrumentally corrected Pre-Search Data Conditioning (PDCSAP) data. By aggregating our multi-iteration ensemble to suppress MCMC stochasticity, we derive a robust global False Positive Probability ($\mathrm{FPP} = 0.00191$) and a Nearby False Positive Probability ($\mathrm{NFPP} < 10^{-6}$), firmly validating the companion well below the rigorous $1.5\%$ threshold established for this Bayesian framework. While the derived $7.8\text{--}8.1\,R_\oplus$ radius places the object at the boundary between giant planets and low-mass brown dwarfs, such a radius is highly atypical for a brown dwarf, pointing to a planetary nature. We report TOI-7701.01 as a validated companion and encourage prompt radial velocity characterization.
\end{abstract}


\section{Introduction} \label{sec:intro}
The Transiting Exoplanet Survey Satellite \citep[TESS;][]{Ricker2015} generates extensive photometric datasets that are highly suitable for automated planet-hunting algorithms. Recently, \citet{Salinas2025} deployed a Transformer-based neural network architecture over TESS Full-Frame Images, successfully detecting a high-significance transit-like signature around the bright F-type subgiant TIC 122522333 ($T_{\rm eff} = 6230$ K, $R_\star = 1.76\,R_\odot$; \citealt{Stassun2019}), designated as TOI-7701.01. While machine-learning frameworks excel at discovering signals in large volumes of data, they do not execute the rigorous statistical testing required to validate a planet's physical veracity, leaving automated alerts vulnerable to astrophysical false positives such as blended eclipsing binaries.

Statistical validation bridges the gap between pipeline detection and formal validation by mathematically ruling out alternative stellar configurations \citep{Torres2011, Giacalone2021}. A critical requirement in this process is the careful management of photometric data conditioning. While Pre-Search Data Conditioning (PDCSAP) arrays optimize the Signal-to-Noise Ratio (SNR) by removing common-mode instrumental systematics and background crowding, they are poorly suited as raw inputs for Bayesian validation engines. These engines model alternative spatial contaminants and thus require unmodified Simple Aperture Photometry (SAP) to evaluate the true photon environment within the aperture.

In this paper, we present the complete independent validation of TOI-7701.01. We leverage multi-sector astrometric tracking and the \texttt{triceratops} Bayesian validation engine—executed through a multi-iteration Markov Chain Monte Carlo (MCMC) framework—to robustly confirm the nature of this sub-Saturn companion. The study details our photometric retrieval in Section \ref{sec:photometry}, examines spatial and centroid configurations in Section \ref{sec:astrometry}, details the Bayesian validation framework in Section \ref{sec:validation}, and discusses the physical implications of the object at the mass-radius boundary in Section \ref{sec:discussion}.

\section{Photometric Reduction and Signal Analysis} \label{sec:photometry}

Our pipeline processes both available data streams from the TESS Science Processing Operations Center (SPOC) to separate geometric characterization from Bayesian false-positive analysis.

\subsection{Geometric Characterization via PDCSAP}
We obtained the light curve of TIC 122522333 observed in Sector 97 from the MAST archive using the \texttt{Lightkurve} infrastructure \citep{Lightkurve2018}. While TESS previously observed this target during Sectors 3, 4, and 30, these historical epochs were captured via Full-Frame Images (FFIs) at longer cadences. Sector 97 provides the only 2-minute short-cadence data processed by the primary Science Processing Operations Center (SPOC) pipeline. We exclusively utilize Sector 97 for the photometric extraction, as the superior temporal resolution (120 seconds) minimizes morphological smearing during transit ingress and egress, ensuring a highly accurate measurement of the transit depth and duration.

A Box Least Squares \citep[BLS;][]{Kovacs2002} periodogram search was executed on the flattened light curve (window length = 1501 cadences), detecting a strong, periodic transit-like signal at $\mathrm{SNR} = 31.49$. The baseline search resolved the following optimal ephemeris: an orbital period of $P = 20.613811$ days, a mid-transit time of $T_0 = 3945.1644$ BTJD, and a total duration of $7.08$ hours.

\begin{figure}[htb!]
    \centerline{\includegraphics[width=0.48\textwidth]{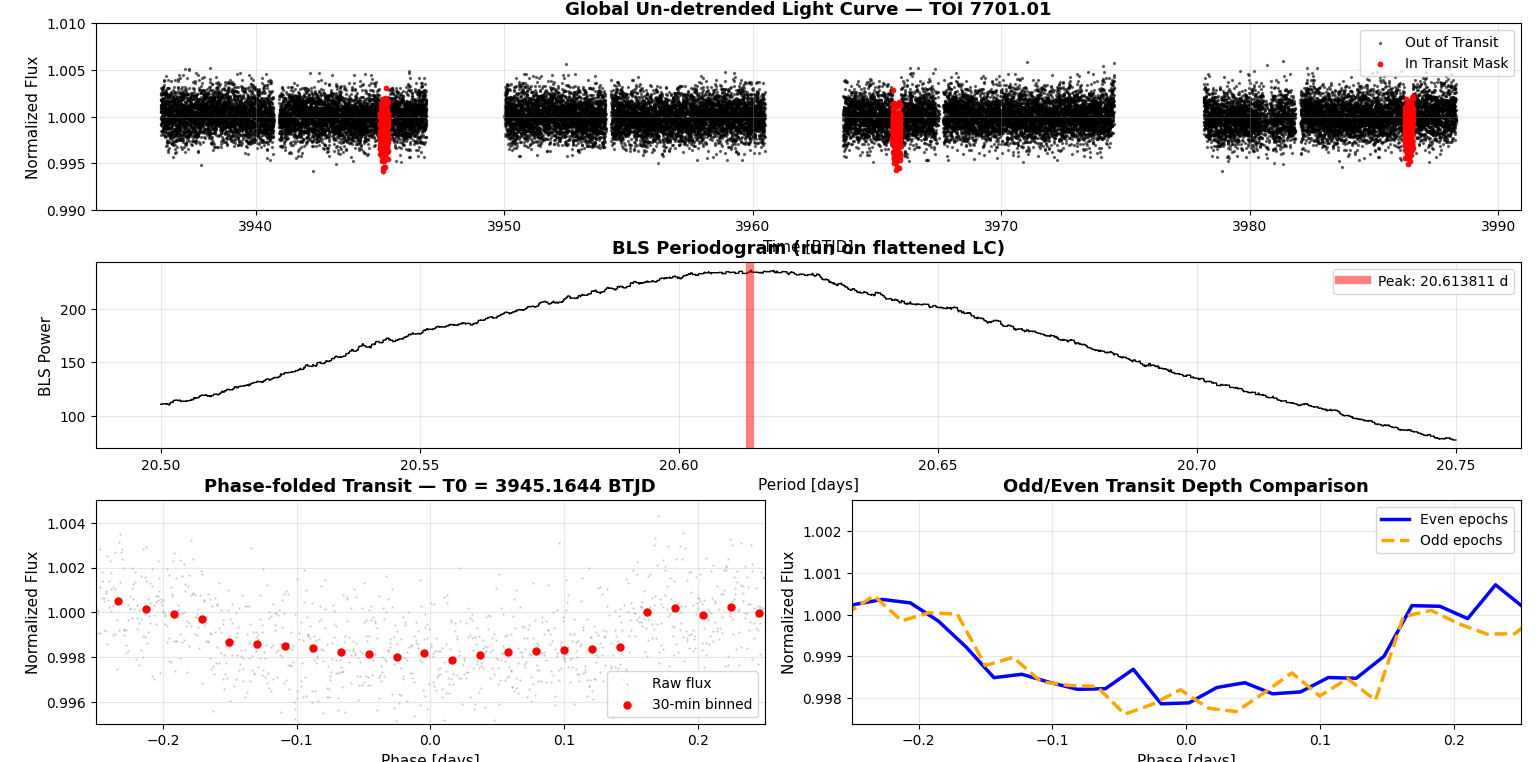}}
    \caption{Photometric transit detection and characterization of TOI-7701.01 using Sector 97 PDCSAP data. \textbf{Top:} The full light curve with in-transit cadences highlighted. \textbf{Middle:} The BLS periodogram showing the primary detection peak at $P \approx 20.61$ days. \textbf{Bottom Left:} The phase-folded transit curve with 30-minute binning, exhibiting a clear morphology with a calibrated depth of $1767$ ppm. \textbf{Bottom Right:} Odd/even transit depth comparison, confirming a uniform primary eclipse and ruling out a near-equal mass eclipsing binary at twice the orbital period.}
    \label{fig:pdcsap}
\end{figure}

The phase-folded transit model demonstrates a well-defined morphology with a calibrated PDCSAP depth of $\delta_{\rm PDCSAP} = 1767$ ppm (Figure \ref{fig:pdcsap}). Because the SPOC processing applies an automated dilution correction factor ($\mathrm{CROWDSAP}$) to account for background starlight contamination, this specific stream represents the true physical occultation depth. Under a standard geometric cross-section assumption ($\delta \approx (R_p/R_\star)^2$) and anchoring the host star at $1.76\,R_\odot$, the companion's radius is derived at $R_p \approx 8.07\,R_\oplus$, characterizing it as a sub-Saturn body.

As a standard false-positive vetting diagnostic, we performed an odd/even transit depth comparison utilizing the instrumentally corrected PDCSAP stream. The phase-folded depths of the alternating odd and even epochs are morphologically consistent and statistically indistinguishable. This consistency confirms the absence of alternating primary and secondary eclipses, effectively ruling out a near-equal mass eclipsing binary scenario at twice the detected orbital period. We explicitly rely on the PDCSAP data for this specific diagnostic, as the un-detrended SAP array retains baseline drifts and stellar variability that could artificially skew comparative depth measurements.

\subsection{Raw Photometric Integration via SAP}
\begin{figure}[htb!]
    \centerline{\includegraphics[width=0.48\textwidth]{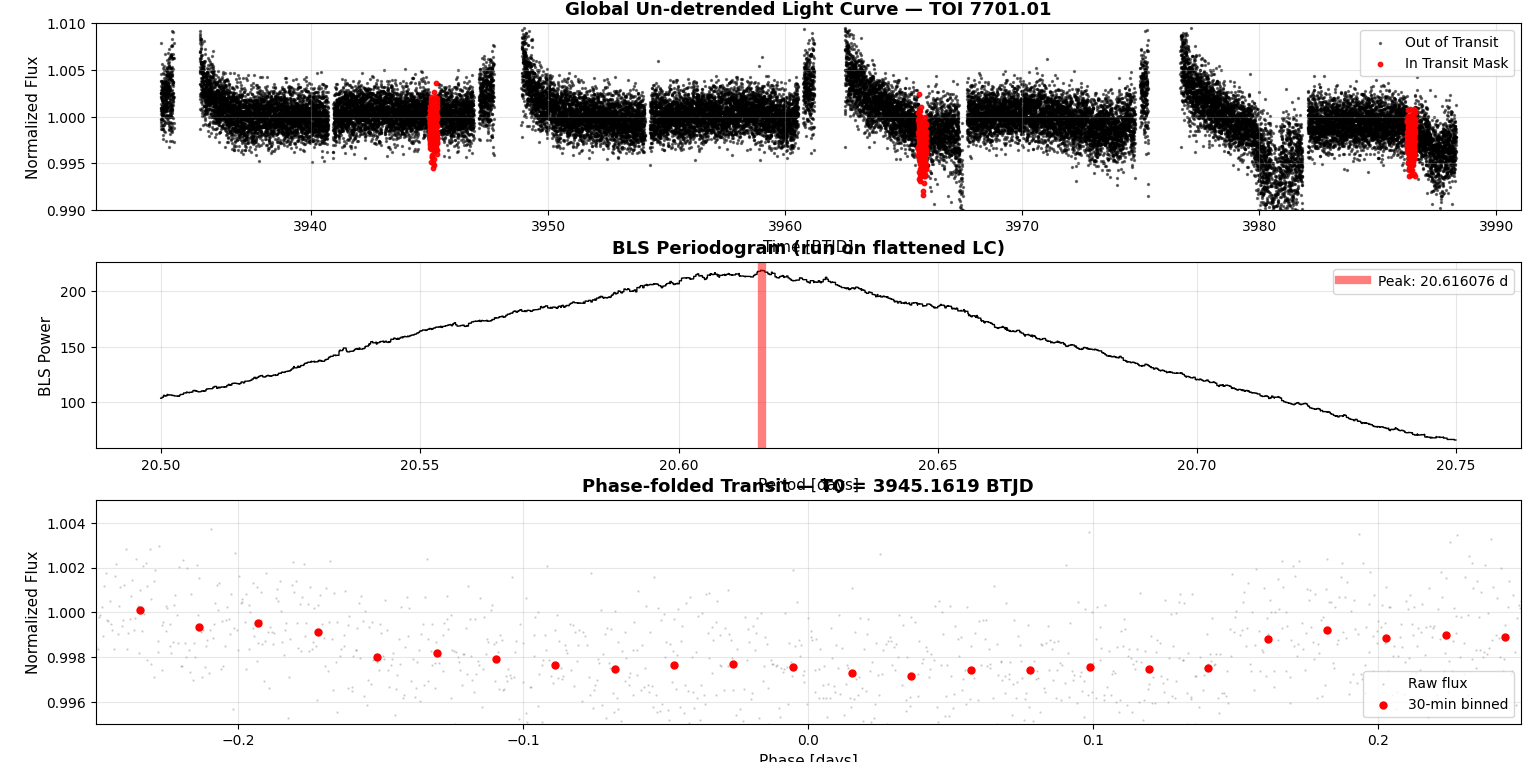}}
    \caption{Photometric signal analysis derived from un-detrended SAP data. \textbf{Top:} The global raw light curve with local out-of-transit baseline normalization. \textbf{Middle:} The BLS periodogram confirming the signal recovery. \textbf{Bottom:} The phase-folded photometry comparing raw flux to 30-minute bins. A zoomed-in view measures a natural, uncorrected transit depth of $2417$ ppm, preserving the original field dilution metrics for downstream validation.}
    \label{fig:sap}
\end{figure}

To supply unmodified data to the validation framework, we extracted the raw SAP flux series. Running the BLS algorithm on a conservatively flattened version of this specific SAP stream yielded a high-significance detection ($\mathrm{SNR} = 26.64$) with an orbital period of $P = 20.616076$ days and an epoch of $T_0 = 3945.1619$ BTJD. Phase-folding the un-detrended SAP data on this exact ephemeris reveals a raw, natural transit depth of $\delta_{\rm SAP} = 2417$ ppm (Figure \ref{fig:sap}). 

Because standard continuous flattening routines (such as Savitzky-Golay filters) are prone to over-subtracting and artificially suppressing deep transit signatures, we conservatively preserved the un-detrended SAP array as the definitive master light curve. To guarantee strict internal data coherence within our pipeline, the specific SAP-derived orbital period ($20.616076$ days) and raw depth ($2417$ ppm) were explicitly utilized as the fixed theoretical inputs for the downstream \texttt{triceratops} statistical engine, applying only a local baseline re-normalization to carefully center the out-of-transit data at unity. A side-by-side comparison of the key photometric parameters derived from both the PDCSAP and SAP pipelines is summarized in Table \ref{tab:photometry}.

\begin{table}[htb!]
    \centering
    \caption{Derived Photometric Parameters for TOI-7701.01}
    \begin{tabular}{lcc}
        \hline \hline
        Parameter & PDCSAP (Geometric) & SAP (Vetting) \\
        \hline
        Orbital Period ($P$) & $20.613811$ days & $20.616076$ days \\
        Epoch ($T_0$) & $3945.1644$ BTJD & $3945.1619$ BTJD \\
        Transit Depth & $1767$ ppm & $2417$ ppm \\
        Signal-to-Noise & $31.49$ & $26.64$ \\
        \hline
    \end{tabular}
    \label{tab:photometry}
\end{table}

\section{Spatial Contamination and Centroid Vetting} \label{sec:astrometry}

Before initiating statistical modeling, we applied spatial and astrometric vetting to confirm that the transit occurs on the target star.

\subsection{Multi-Sector Astrometric Contamination}
We executed a spatial query using Gaia DR3 \citep{Gaia2023} within a 1-arcminute radius centered on the target ($G = 11.08$). To account for stellar kinematics over multi-year baselines, the proper motion vectors were propagated from the Gaia J2016.0 epoch to the exact epoch of each TESS observation across all historical sectors (Sectors 3, 4, 30, and 97). 

\begin{figure}[htb!]
    \centerline{\includegraphics[width=0.48\textwidth]{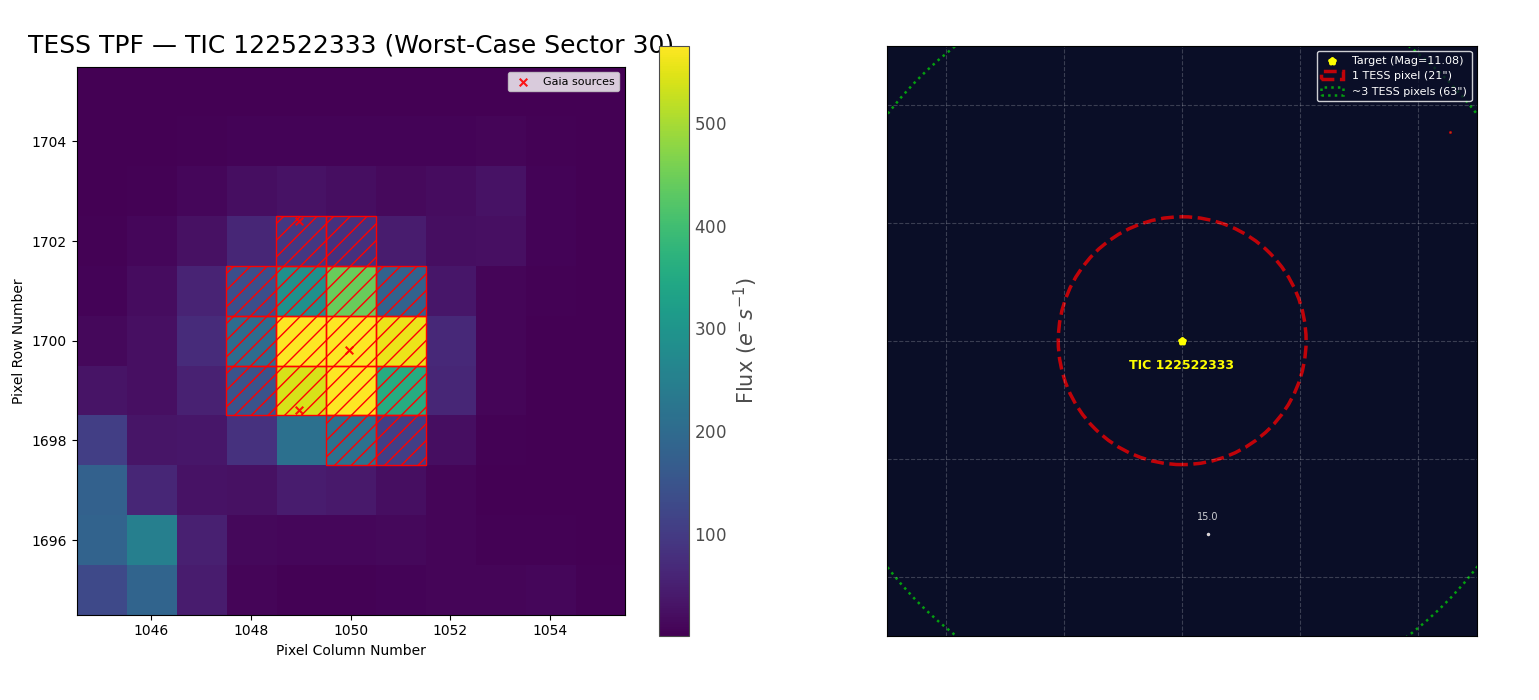}}
    \caption{Spatial contamination map based on Gaia DR3 coordinates propagated to the TESS epoch. Sources with poorly constrained solutions ($\mathrm{RUWE} > 1.4$) are marked with red borders.}
    \label{fig:gaia_map}
\end{figure}

Our pipeline mapped the target pixel response function against the designated extraction apertures for each sector. The analysis identified the worst-case background contamination scenario in Sector 30, limiting the maximum flux dilution inside the aperture to $3.00\%$ (Figure \ref{fig:gaia_map}). This dilution originates entirely from two faint background sources: a primary contaminant ($G = 15.01$) at a separation of $33.0^{\prime\prime}$, and a secondary source ($G = 17.31$) at $57.6^{\prime\prime}$. 

Crucially, the secondary source exhibits a highly anomalous Renormalised Unit Weight Error ($\mathrm{RUWE} = 3.94$). A $\mathrm{RUWE} > 1.4$ indicates a poorly constrained astrometric solution, strongly symptomatic of an unresolved binary system. Because this source is $\Delta G = 6.23$ magnitudes fainter than the target, it contributes a mere $\approx 0.32\%$ ($\sim 3200$ ppm) to the total aperture flux. For this specific star to replicate our observed $2417$ ppm transit signal, it would need to undergo a massive, near-total eclipsing event ($\approx 75\%$ depth). While this establishes a theoretical Background Eclipsing Binary (BEB) scenario that requires explicit vetting, its large angular separation (nearly three $21^{\prime\prime}$ TESS pixels) implies that such a deep localized eclipse would induce a severe and readily detectable shift in the photometric centroid.

\subsection{Sub-pixel Centroid Tracking}
To determine if the transit signal originated from the primary target or a nearby contaminant, we tracked the sub-pixel centroid position during the Sector 97 transit window. We computed flux-weighted moments on the native Target Pixel File (TPF) and applied a high-pass 631-cadence (21-hour) median filter to isolate spacecraft jitter from astrophysical shifts. 

\begin{figure}[htb!]
    \centerline{\includegraphics[width=0.48\textwidth]{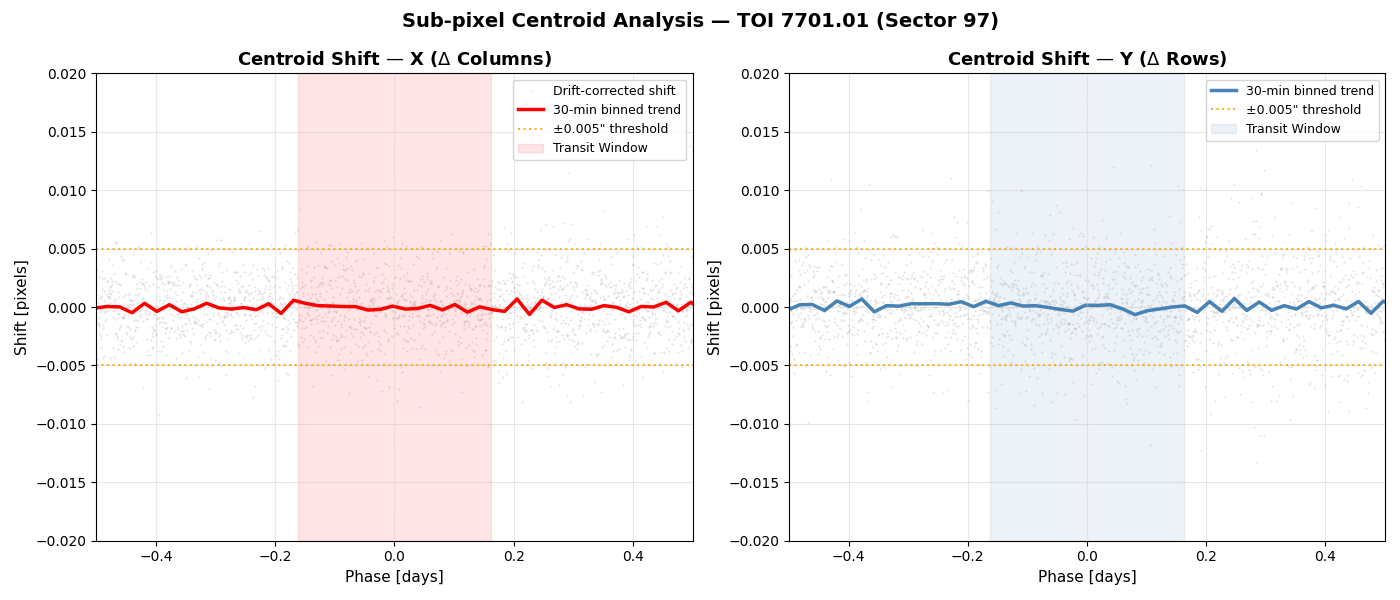}}
    \caption{Phase-folded X and Y centroid displacements during the Sector 97 transit window. The residual tracking remains tightly bounded near zero, confirming on-target stability.}
    \label{fig:centroids}
\end{figure}

The phase-folded residual centroid series reveals exceptional spatial stability, with maximum in-transit displacements restricted to $\Delta X = 0.0004$ pixels and $\Delta Y = 0.0007$ pixels (Figure \ref{fig:centroids}). Both dimensions remain well inside the conservative tracking threshold of $\pm0.005$ pixels. This measurement confirms that the transit-like dimming is centered on the primary target coordinates, effectively ruling out the high-RUWE background star and other surrounding sources as the origin of the signal.

\section{Statistical Validation Framework} \label{sec:validation}

We employed the Bayesian validation code \texttt{triceratops} \citep{Giacalone2021} to compute probabilities across fifteen distinct target star configurations and three neighboring star scenarios.

\subsection{Ensemble Calibration and Parameter Convergence}
Raw SAP light curves often present subtle baseline features that clash with the flat out-of-transit profile assumed by transit engines. To prevent data clipping or baseline mismatches from skewing the results, we isolated the out-of-transit photometry ($|t| > 0.20$ days) and re-normalized the entire array by its median, centering the baseline at exactly $1.000$. Furthermore, because single MCMC implementations are susceptible to localized chain stagnation, we executed a robust stability check consisting of a 20-iteration ensemble run to compute mean probabilities and variances.

A major result of this configuration is the structural convergence of the derived physical radius. Even though the engine was fed the raw un-detrended SAP light curve (reflecting a deep $2417$ ppm profile), the application of the built-in \texttt{triceratops} prior distributions and stellar models automatically constrained the dominant Target Planet (TP) scenario ($\mathcal{P}_{\rm TP} = 78.1\%$) to an adjusted radius of $R_p = 7.86\,R_\oplus$, evaluated on a single reference iteration, while other variations within the ensemble have shown maximum values of $\approx 8.1\,R_\oplus$. This parameter convergence demonstrates an elegant agreement with the independent geometric radius ($8.07\,R_\oplus$) derived from the background-corrected PDCSAP stream, placing the companion within a narrow $7.8\text{--}8.1\,R_\oplus$ window.

\begin{figure}[htb!]
    \centerline{\includegraphics[width=0.48\textwidth]{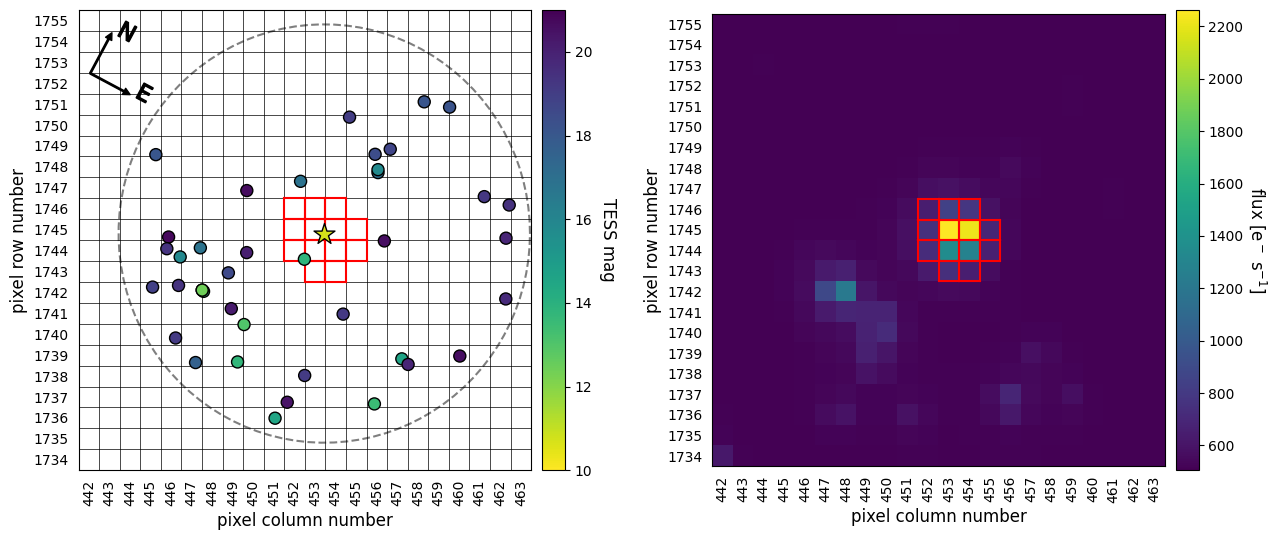}}
    \caption{TESS SPOC pixel aperture configuration for Sector 97 showing nearby field sources assessed by the \texttt{triceratops} Bayesian matrix.}
    \label{fig:triceratops_out}
\end{figure}

\subsection{Bayesian Vetting Metrics}
The multi-iteration MCMC ensemble converged on a global False Positive Probability of $\mathrm{FPP} = 0.00191 \pm 0.00247$ and a Nearby False Positive Probability of $\mathrm{NFPP} < 10^{-6}$ (numerically consistent with zero). Both values satisfy the validation criteria ($\mathrm{FPP} < 0.015$, $\mathrm{NFPP} < 0.001$) defined by \citet{Giacalone2021}, meaning the planetary scenario is favored over any alternative configuration. 

To examine the detailed probability breakdown across individual astrophysical models, we evaluated a single, high-convergence reference iteration representative of the ensemble mean. In this reference distribution, the Bayesian engine allocates a combined $99.7\%$ likelihood to configurations containing a genuine transiting planet. This encompasses:
\begin{itemize}
    \item \textbf{Target Planet (TP):} $78.1\%$. The system contains no unresolved companions, and the transiting planet orbits the target star.
    \item \textbf{Primary Transiting Planet (PTP):} $14.8\%$. The system contains an unresolved bound stellar companion, but the transiting planet still orbits the primary target star.
    \item \textbf{Diluted Target Planet (DTP):} $6.8\%$. The aperture contains an unresolved background star, but the transiting planet orbits the target star.
\end{itemize}

\begin{figure}[htb!]
    \centerline{\includegraphics[width=0.48\textwidth]{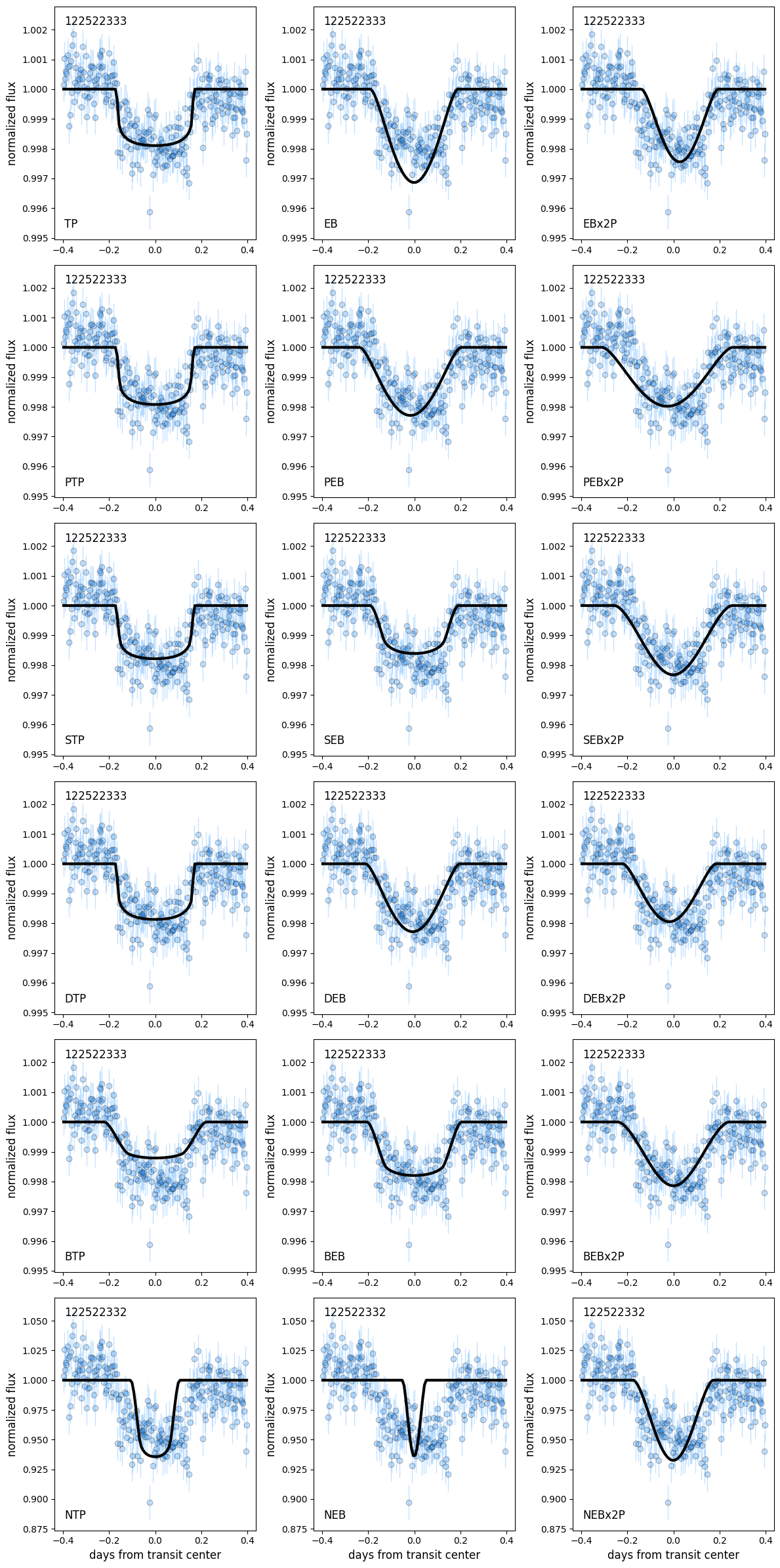}}
    \caption{Astrophysical transit models generated by \texttt{triceratops} fitted against the re-normalized SAP curve. Planetary models optimize the fit, while binary models are statistically excluded.}
    \label{fig:scenarios}
\end{figure}

Crucially, under all three of these dominant probabilistic models, the transit originates from a planetary-mass companion orbiting the primary target star (TIC 122522333). Conversely, all eclipsing binary models (whether on the target, bound companions, or background sources) are heavily penalized in the posterior distribution (see Figure \ref{fig:scenarios}), driven down by the clean U-shaped transit morphology and the lack of centroid movement.

\section{Discussion: The Giant Planet Boundary} \label{sec:discussion}
The convergence of our geometric and Bayesian pipelines establishes a consistent radius estimate within the $7.8\text{--}8.1\,R_\oplus$ range. This scale places TOI-7701.01 at a challenging physical boundary where transit photometry alone faces inherent mass-radius degeneracies. Structurally, an object of this size could theoretically describe either a sub-Saturn gas giant or a very compact, low-mass brown dwarf. However, because electron degeneracy pressure typically causes brown dwarfs to maintain radii closer to $\approx 1\,R_{\rm Jup}$ ($\sim 11.2\,R_\oplus$) across a wide range of masses \citep{Chabrier2009}, a radius of $\sim 8\,R_\oplus$ is highly atypical for a brown dwarf, strongly favoring a planetary internal structure.

To definitively break this degeneracy in the absence of radial velocity mass measurements, we must weigh the astronomical priors of the two populations. With an orbital period of $P \approx 20$ days, the companion resides squarely within the "brown dwarf desert" \citep{Grether2006}—a well-documented region of parameter space where brown dwarf companions to solar-type and moderately massive stars are statistically exceedingly rare compared to gas giants. The combination of its sub-Saturn radius, the strict population priors, and our rigorously low $\mathrm{FPP}$ ($0.191\%$) clearly supports the planetary scenario. Furthermore, its presence around an evolved F-type subgiant ($R_\star = 1.76\,R_\odot$) makes it a valuable data point for studying the formation and potential re-inflation mechanisms of sub-Saturns as their host stars leave the main sequence.

\section{Conclusion} \label{sec:conclusion}
We have presented the formal statistical validation of TOI-7701.01, successfully elevating the target from an automated machine-learning detection to a comprehensively validated companion. By conducting a dual photometric analysis—leveraging un-detrended SAP data for unbiased spatial false-positive evaluation and PDCSAP data for geometric characterization—we demonstrated robust parameter convergence. Backed by highly stable centroid tracking and a multi-iteration \texttt{triceratops} ensemble yielding an $\mathrm{FPP}$ of $0.191\%$, the presence of the transiting companion is firmly established, and a planetary classification is highly favored.

Given the brightness of the subgiant host star ($V = 11.09$) and the companion's critical position at the giant planet boundary, TOI-7701.01 emerges as a great laboratory for follow-up studies. We strongly encourage precision radial velocity (PRV) observations to definitively rule out a low-mass brown dwarf alternative and constrain the object's absolute mass and bulk density. Establishing these parameters will resolve its core mass fraction and determine whether it formed via core accretion or disk instability, ultimately contributing to our broader understanding of the sub-Saturn population.

\section*{Acknowledgments}
This paper includes data collected by the TESS mission, which are publicly available from the Mikulski Archive for Space Telescopes (MAST). Funding for the TESS mission is provided by NASA's Science Mission Directorate. We acknowledge the use of public data from the Exoplanet Follow-up Observation Program (ExoFOP), which is operated by the California Institute of Technology, under contract with the National Aeronautics and Space Administration. 

This work has made use of data from the European Space Agency (ESA) mission \textit{Gaia} (\url{https://www.cosmos.esa.int/gaia}), processed by the \textit{Gaia} Data Processing and Analysis Consortium (DPAC). Funding for the DPAC has been provided by national institutions, in particular the institutions participating in the \textit{Gaia} Multilateral Agreement. This research has made use of the VizieR catalogue access tool, CDS, Strasbourg, France.

\facilities{TESS (MAST), Gaia}

\software{
    Lightkurve \citep{Lightkurve2018},
    Triceratops \citep{Giacalone2021},
    Astropy \citep{Astropy2022},
    Astroquery \citep{Ginsburg2019},
    Matplotlib \citep{Hunter2007},
    Pandas, NumPy, SciPy.
}

\end{document}